\documentclass[conference,10pt]{IEEEtran}
\usepackage{epsfig,rotating,setspace,latexsym,amsmath,epsf,amssymb,amsfonts,bm,theorem,subfigure,epstopdf}
\usepackage{cite,authblk}
\usepackage{bbm}

\usepackage{algorithm}
\usepackage[noend]{algpseudocode}

\algrenewcommand\algorithmicforall{\textbf{foreach}}
\algrenewcommand\algorithmicindent{.8em}

\usepackage{color}
\usepackage{mathtools}

\newtheorem{theorem}{Theorem}

\newenvironment{Proof}[1]{\medskip\par\noindent{\bf Proof:\,}\,#1}{{\mbox{\,$\blacksquare$}\par}}
\IEEEoverridecommandlockouts
\allowdisplaybreaks

\begin{document} 
	\title{Gossiping with Binary Freshness Metric \thanks{This work was supported by NSF Grants CCF 17-13977 and ECCS 18-07348.}}
    \author{Melih Bastopcu  \qquad Baturalp Buyukates \qquad Sennur Ulukus\\
	\normalsize Department of Electrical and Computer Engineering\\
	\normalsize University of Maryland, College Park, MD 20742\\
	\normalsize \emph{bastopcu@umd.edu}  \qquad \emph{baturalp@umd.edu} \qquad  \emph{ulukus@umd.edu}}
	
\maketitle

\begin{abstract}
We consider the binary freshness metric for gossip networks that consist of a single source and $n$ end-nodes, where the end-nodes are allowed to share their stored versions of the source information with the other nodes. We develop recursive equations that characterize binary freshness in arbitrarily connected gossip networks using the stochastic hybrid systems (SHS) approach. Next, we study binary freshness in several structured gossip networks, namely disconnected, ring and fully connected networks. We show that for both disconnected and ring network topologies, when the number of nodes gets large, the binary freshness of a node decreases down to 0 as $n^{-1}$, but the freshness is strictly larger for the ring topology. We also show that for the fully connected topology, the rate of decrease to 0 is slower, and it takes the form of $n^{-\rho}$ for a $\rho$ smaller than 1, when the update rates of the source and the end-nodes are sufficiently large. Finally, we study the binary freshness metric for clustered gossip networks, where multiple clusters of structured gossip networks are connected to the source node through designated access nodes, i.e., cluster heads. We characterize the binary freshness in such networks and numerically study how the optimal cluster sizes change with respect to the update rates in the system.   %with a strictly larger multiplier for the ring topology.  
\end{abstract}
 
\section{Introduction}\label{sect:intro}
Timeliness of the received information has gained significant attention with the emergence of modern technologies such as autonomous driving, holographic communications, and industrial IoT. To measure the timeliness at the receiver, age of information (AoI) has been proposed as a performance metric \cite{Kaul12a}; see also \cite{YatesSurvey} for a recent survey. The traditional age metric increases linearly over time in the absence of any update deliveries and reduces to the age of the most recently received  update upon successful delivery. In its original form, the age metric does not encapsulate the information change rate at the source. For example, when there has been no update deliveries, the age at the receiver gets large indicating that the information at the receiver has become stale. On the contrary, the receiver may still have the freshest version of the information even though it has not received any updates for a long time, if the information at the source does not change frequently. 

Motivated by this, several variations of the traditional age metric have been proposed recently to consider information change rate at the source. In this work, we use the binary freshness metric to characterize information freshness at the receiver nodes. Previously used in \cite{cho03, kolobov19a, Bastopcu2021, Bastopcu20e, Kaswan21}, binary freshness takes the value $1$ when the information at the receiver node is up-to-date whereas it takes the value $0$ when the receiver node has a stale information. This metric is similar to version age of information \cite{Bastopcu20c, Yates21, Buyukates21c, Eryilmaz21} in that both metrics remain unchanged as long as the information at the source stays the same. While version age increases by $1$ whenever the source generates a new update, binary freshness immediately becomes $0$ as soon as the receiver node and the source are out-of-sync. Another metric that considers the information change statistics at the source is age of incorrect information (AoII), which is equal to $0$ if the receiver has fresh information, otherwise, it is larger than $0$ and may increase over time \cite{Maatouk20b, Maatouk20a, Kam20a}. A special case of AoII is the age of synchronization introduced in \cite{Zhong18c}.

\begin{figure}[t]
	\centerline{\includegraphics[width=0.8\columnwidth]{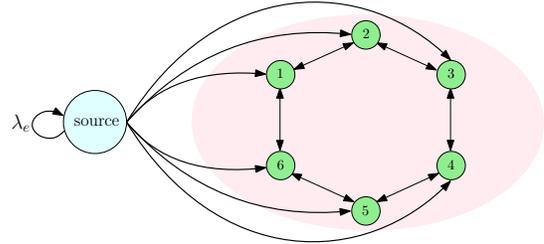}}
	\vspace{-0.2cm}
	\caption{Gossip network model consisting of a source represented by the blue node, and the users represented by the green nodes. Here, users form a ring network. Other network topologies are shown in Fig.~\ref{Fig:netw_types}.}
	\vspace{-0.5cm}
	\label{Fig:disconnected}
\end{figure}

Our aim in this work is to study information timeliness in arbitrarily connected and structured gossip networks using the binary freshness metric. Works that are closely related to our work are \cite{Bastopcu2021, Bastopcu20e, Kaswan21, Yates21, Buyukates21c}. The binary freshness metric has been studied for serially connected caching systems in \cite{Bastopcu2021, Bastopcu20e} and for parallel connected caching systems in \cite{Kaswan21}. References \cite{Bastopcu2021, Bastopcu20e, Kaswan21} maximize information freshness at the end-nodes in caching systems by using alternating maximization methods. Different from \cite{Bastopcu2021, Bastopcu20e, Kaswan21}, in this work, we employ the stochastic hybrid systems (SHS) approach and develop a general method that enable us to characterize the binary freshness in \emph{arbitrarily} connected networks. We explore gossiping strategies among the end-nodes and study freshness in structured gossip networks, i.e., disconnected, ring and fully connected network topologies. Reference \cite{Yates21} uses the SHS approach to characterize the version age and finds the scaling of version age in structured gossip networks. Reference \cite{Buyukates21c} improves the scaling of version age by introducing \emph{clustering} to gossip networks. Here, we develop the binary freshness counterpart of the works \cite{Yates21, Buyukates21c}. As the binary freshness metric and encompassing mathematics are different than those in version age, our work is distinct from works \cite{Yates21, Buyukates21c}.

\begin{figure}[t]
	\centerline{\includegraphics[width=0.75\columnwidth]{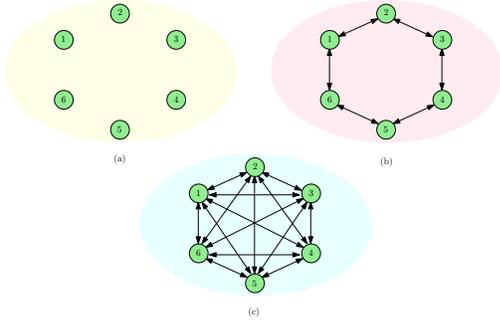}}
	\vspace{-0.2cm}
	\caption{(a) Disconnected, (b) ring, and (c) fully connected network topologies with $n=6$ end-nodes.}
	\label{Fig:netw_types}
	\vspace{-0.5cm}
\end{figure}

In this work, by using the SHS approach, we first provide a way to characterize binary freshness in arbitrarily connected networks. Then, we consider binary freshness in structured gossip networks, such as disconnected, ring and fully connected networks; see Figs.~\ref{Fig:disconnected}-\ref{Fig:netw_types}. We show that when the number of nodes becomes large, the binary freshness of a node decreases down to $0$ with rate $n^{-1}$ for the disconnected and ring networks, but the freshness is strictly larger for the ring networks. For the fully connected networks, when the update rates of the source and the end-nodes are sufficiently large, the binary freshness of a node decreases down to $0$ slower than the other network types, indicating that increased connectivity among the nodes improves the freshness. Finally, we find the binary freshness in clustered gossip networks, where each cluster consists of a structured gossip network and is connected to the source through a designated access node, i.e., a cluster head. We numerically find the optimal number of nodes in each cluster for given update rates of the source, cluster heads, and the end-nodes.   

\section{Freshness in Arbitrarily Connected Networks}\label{sect:freshness}
In this section, we consider freshness in gossip networks with arbitrary topologies. System model considered in this section is complementary to the models in \cite{Yates21, Buyukates21c}, in that we use the binary freshness metric to characterize timeliness in gossip networks as opposed to the version age metric used in \cite{Yates21, Buyukates21c}. In our model, there is a single source node that is updated as a Poisson process with rate $\lambda_e$. This source node updates the nodes in the system with a total rate of $\lambda$. There are $n$ nodes in the system, which are denoted by the set $\mathcal{N} \triangleq \{1,2,\ldots, n\}$. Nodes in the system implement gossiping as in \cite{Yates21, Buyukates21c} to distribute update packets among each other. The total update rate of a node is $\lambda$, and each node in $\mathcal{N}$ uniformly distributes its update rate among its neighbors. 

We first characterize the binary freshness in arbitrarily connected gossip networks. Following the analysis for the version age in \cite{Yates21, Buyukates21c}, we write the freshness $F_i$ in terms of $F_S$, which denotes the freshness of a subset $S$ and it is given by $F_S(t) \triangleq \max_{j \in S} F_j (t)$. We note that here we have a maximum operator as opposed to a minimum operator since our freshness metric takes the maximum value when the information at the receiver node is the freshest. 

As in \cite{Yates21}, we denote the total update rate from node $i$ into set $S$ by $\lambda_i(S)$, i.e., $\lambda_i (S) = \sum_{j \in S} \lambda_{ij}$ when $i \notin S$. Similarly, $\lambda_s(S)$ denotes the total update rate from the source into set $S$. Set $N(S)$ denotes the set of updating neighbors of $S$ given by $N(S) = \{ i \in \{1,\ldots,n \} : \lambda_i (S) > 0 \}$. 

In Theorem~\ref{freshness_gossip}, we characterize the freshness in arbitrarily connected gossip networks. This result is the binary freshness equivalent of the version age result in \cite[Thm.~1]{Yates21}.

\begin{theorem}\label{freshness_gossip}
  The average freshness of a set $S$, $F_S = \lim_{t \to \infty} \mathbb{E}[F_S(t)]$, for $S \subseteq \mathcal{N}$ is given by
  \begin{align}\label{freshness_gossip_res}
      F_S = \frac{\lambda_s(S) + \sum_{i\in N(S)} \lambda_i{(S)} F_{S\cup\{i\}}} {\lambda_e + \lambda_s(S) + \sum_{i\in N(S)}\lambda_i(S)}.
  \end{align}
\end{theorem}

\begin{Proof}
A state transition in the system happens when a node $i$ updates node $j$. We first present the possible state transitions. Let $\mathcal{L}$ denote the set of transitions as in \cite{Yates21}. Accordingly, 
\begin{align}\label{transitions}
    \mathcal{L} = \{ (s,s)\} \cup \{ (s,j): j\in \mathcal{N}\} \cup \{(i,j): i,j \in \mathcal{N} \},
\end{align}
where the first transition occurs when the source generates a new update, the second set of transitions occur when the source node directly updates an end-node in $\mathcal{N}$, and the third set of transitions occur when an end-node updates another end-node. In our case, different than \cite{Yates21}, the freshness vector evolves as
\begin{align}\label{age_transitions}
    F'_k = \begin{cases} 
      0, & i=j=s, k\in\mathcal{N}, \\
      1, & i=s, j=k\in\mathcal{N}, \\
      \max(F_i, F_j), & i\in\mathcal{N}, j=k\in\mathcal{N},\\
      F_k, & \text{otherwise},
   \end{cases} 
\end{align}
where $F'_k$ is the freshness of node $k$ after a transition. In (\ref{age_transitions}), freshness takes the minimum value of $0$ when node $k$ has stale information while it takes the maximum value of $1$ when node $k$ has fresh information, i.e., the same information as the source. After the $(s,s)$ transition, the freshness of the set $S$ becomes $0$ as all the nodes in set $S$ become out-of-date. For other transitions $(i,j)$, given that $j\in S$, we have
\begin{align}\label{freshness_update}
    F'_S = \max_{k\in S} F'_k = \max_{k \in S\cup\{i\}}F_k = F_{S\cup \{i\}}.
\end{align}
We note that when $i=s$, (\ref{freshness_update}) implies that $F'_S = 1$. If $j\notin S$, the freshness of set $S$ is unchanged after transition $(i,j)$, i.e., $F'_S = F_S$. Using (\ref{age_transitions}) and (\ref{freshness_update}) and following similar steps as in \cite{Yates21} yields the result.
\end{Proof}

\section{Sample Freshness Evaluations}\label{sect:examples}
To demonstrate the application of (\ref{freshness_gossip_res}), in what follows, we consider simple network examples and characterize the average binary freshness experienced by the end-nodes. 

\begin{figure}[t]
 	\begin{center}
 	\subfigure[]{%
 	\includegraphics[width=0.49\linewidth]{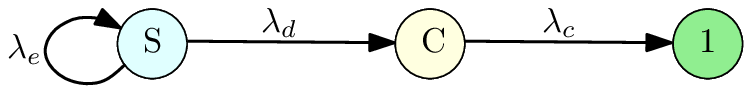}}\\ \vspace{-0.2cm}
 	\subfigure[]{%
 	\includegraphics[width=0.49\linewidth]{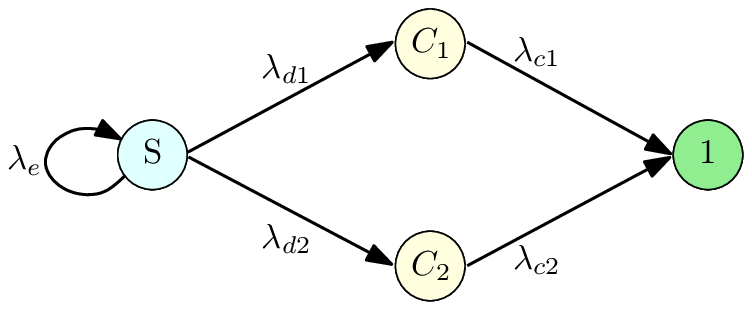}}
 	\subfigure[]{%
 	\includegraphics[width=0.49\linewidth]{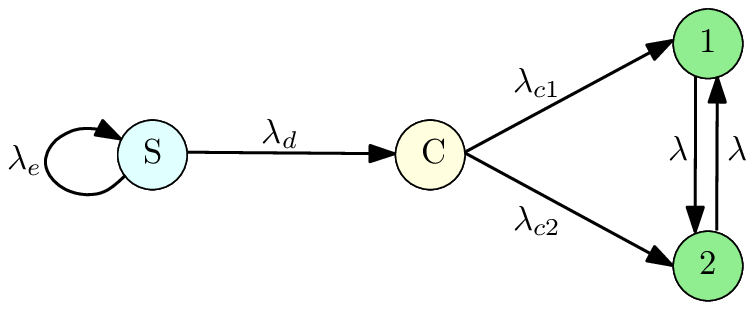}}
 	\end{center}
 	\vspace{-0.4cm}
 	\caption{Freshness of information in (a) a serially connected network, (b) a parallel connected network, and (c) an arbitrarily connected network.}
 	\label{Fig:examples}
 	\vspace{-0.5cm}
\end{figure}

First, we consider the network in Fig.~\ref{Fig:examples}(a). The information at the source is updated with rate $\lambda_e$. The source sends updates to the cache with rate $\lambda_d$ and the cache sends updates to node $1$ with rate $\lambda_c$. By noting that only the cache updates node $1$ and there is no direct link from the source to node $1$, we have $N(1) = C$ and $\lambda_s(1) = 0$. Thus, from (\ref{freshness_gossip_res}), we find $F_1 = \frac{\lambda_c F_{\{1\cup C\}}}{\lambda_e+\lambda_c}$. As the subset $\{1\cup C\}$ gets updates only from the source, we have $N(\{1\cup C\}) = \emptyset$ and $ \lambda_s(\{1\cup C\}) = \lambda_d$. Thus, $F_{\{1\cup C\}}$ becomes $\frac{\lambda_d}{\lambda_d+\lambda_e}$. By inserting $F_{\{1\cup C\}}$ into $F_1$, 
\begin{align}\label{example1}
    F_1 = \frac{\lambda_c}{\lambda_c+\lambda_e}\frac{\lambda_d}{\lambda_d+\lambda_e}. 
\end{align}
Note that, in this example, the nodes are connected serially and the freshness expression in (\ref{example1}) is the same as \cite[(11)]{Bastopcu2021}.

Second, we consider the network in Fig.~\ref{Fig:examples}(b) where we have a source, two caches connected in parallel, and a user. We want to find the freshness at node 1, i.e., $F_1$. Since $N(1) = \{C_1, C_2\}$ and $\lambda_s(1)= 0$, $F_1$ is equal to
\begin{align}\label{example3_int}
    F_1 = \frac{\lambda_{c1}F_{\{1\cup C_1\}}+\lambda_{c2} F_{\{1\cup C_2\}}}{\lambda_e+\lambda_{c1}+\lambda_{c2}}.
\end{align}
Then, we find $F_{\{1\cup C_1\}} = (\lambda_{d1} + \lambda_{c2}F_{\{1\cup C_1\cup C_2\}})/(\lambda_e+\lambda_{d1}+\lambda_{c2}),$    $F_{\{1\cup C_2\}} =  (\lambda_{d2} + \lambda_{c1}F_{\{1\cup C_1\cup C_2\}})/(\lambda_e+\lambda_{d2}+\lambda_{c1}),$  $F_{\{1\cup C_1\cup C_2\}} = (\lambda_{d1} + \lambda_{d2})/(\lambda_e+\lambda_{d1}+\lambda_{d2})$. Inserting $F_{\{1\cup C_1\}}$, $F_{\{1\cup C_2\}}$, and $F_{\{1\cup C_1\cup C_2\}}$ back into (\ref{example3_int}), we obtain 
\begin{align}
    F_1 = &\frac{(\lambda_{d1} + \lambda_{d2})(\lambda_{c1} + \lambda_{c2})}{(\lambda_e+\lambda_{d1} + \lambda_{d2})(\lambda_e+\lambda_{c1} + \lambda_{c2})} \nonumber\\
    &-\frac{\lambda_e}{(\lambda_e+\lambda_{d1} + \lambda_{d2})(\lambda_e+\lambda_{c1} + \lambda_{c2})} \times \nonumber\\
    & \quad \left(\frac{\lambda_{c2}\lambda_{d1}}{\lambda_e+\lambda_{c1}+\lambda_{d2}}+\frac{\lambda_{c1}\lambda_{d2}}{\lambda_e+\lambda_{c2}+\lambda_{d1}}\right). \label{example3_final}
\end{align}
Note that, in this example, the caches are connected in parallel and the freshness expression in (\ref{example3_final}) is the same as \cite[(21)]{Kaswan21} but its derivation is much simpler here.

Third, we consider the network in Fig.~\ref{Fig:examples}(c). In this example, we want to find the freshness at node 1, i.e., $F_1$. Since $N(1) = \{2, C\}$ and $\lambda_s(1)= 0$, $F_1$ is equal to
\begin{align}\label{example2_int}
    F_1 = \frac{\lambda_{c1}F_{\{1\cup C\}}+\lambda F_{\{1\cup2\}}}{\lambda_e+\lambda_{c1}+\lambda}.
\end{align}
Similarly, we find $F_{\{1\cup C\}}$, $F_{\{1\cup 2\}}$, and $F_{\{1\cup2\cup C\}}$. By using these in (\ref{example2_int}), we obtain 
\begin{align}\label{example2}
    F_1 = & \frac{\lambda_{c1}}{\lambda_e+\lambda_{c1}+\lambda} \frac{\lambda_d +\lambda \frac{\lambda_d}{\lambda_d+\lambda_e} }{\lambda_e+\lambda_d+\lambda}\nonumber\\
    &+ \frac{\lambda}{\lambda_e+\lambda_{c1}+\lambda} \frac{(\lambda_{c1}+\lambda_{c2})\frac{\lambda_d}{\lambda_d+\lambda_e}}{\lambda_e+\lambda_{c1}+\lambda_{c2}}. 
\end{align}
Note that, in this example, the nodes are arbitrarily connected, i.e., the network includes both serial and parallel connections. Prior to this work, the freshness expression was known only for serially connected networks \cite{Bastopcu2021} and parallel relay networks \cite{Kaswan21}. Thus, with the method developed here, we are able to find the freshness expression for arbitrarily connected networks.   

\section{Freshness in Structured Gossip Networks}\label{sect:gossip}
In this section, we consider information freshness in structured networks. We consider three different network structures regarding the connectivity among the end-nodes: disconnected, ring and fully connected networks. In all cases, there is a single source node and $n$ end-nodes. The information at the source node is updated with rate $\lambda_e$. The source node updates each end-node with rate $\frac{\lambda}{n}$. Each end-node updates its immediate neighbors with a total  $\lambda$ update rate equally distributed over the neighbors. In what follows, we denote $\rho \triangleq \frac{\lambda_e}{\lambda}$. 

\subsection{Disconnected Networks}
In this network, the end-nodes are not connected to each other and only the source node updates the end-nodes with rate $\frac{\lambda}{n}$ for each node; this network is obtained when Fig.~\ref{Fig:netw_types}(a) is inserted into Fig.~\ref{Fig:disconnected}. Thus, for an arbitrary node $S_1$, we have $\lambda_s(S_1) = \frac{\lambda}{n}$ and $N(S_1) = \emptyset$. Then, from (\ref{freshness_gossip_res}), we have
\begin{align}\label{freshness_disconnected_gossip}
    F_{S_1}  = \frac{1}{1+n\rho}. 
\end{align}
Hence, the freshness of a node goes to 0 as $\frac{1}{n}$ with the network size $n$. This is stated formally in Theorem~\ref{disconnected_thm} below.

\begin{theorem}\label{disconnected_thm}
  In a disconnected network, the average freshness of a single node decreases down to $0$ as $\frac{1}{\rho} n^{-1}$.
\end{theorem}

\subsection{Ring Networks}
In this network, the end-nodes are connected to each other as a bidirectional ring where each node updates its two neighbors with rate $\frac{\lambda}{2}$ each, and the source node updates the end-nodes with rate $\frac{\lambda}{n}$ for each node; this is the network in Fig.~\ref{Fig:disconnected}. Here, subset $S_j$ denotes any arbitrary $j$ adjacent nodes. In particular, if $S_j= \{1,\dots,j\}$, then we have $\lambda_s(S_j) = \frac{j\lambda}{n}$, and $N(S_j) = \{j+1,n\}$ (see Fig.~\ref{Fig:netw_types}(b)). Thus, from (\ref{freshness_gossip_res}), we have 
\begin{align}\label{ring_recursion}
    F_{S_j} = \frac{\frac{j\lambda}{n}+\lambda F_{S_{j+1}}}{\lambda_e +\lambda+\frac{j\lambda}{n} },
\end{align}
for $j =1, \dots, n-1$, and $F_{S_n} = \frac{1}{1+ \rho}$. Theorem~\ref{ring_thm} below states the form of the freshness of a node for large $n$. We note that the freshness in a ring network in Theorem~\ref{ring_thm} is larger than the freshness in a disconnected network in Theorem~\ref{disconnected_thm}.

\begin{theorem}\label{ring_thm}
  In a ring network, the average freshness of a single node decreases down to $0$ as $\left(\frac{1}{\rho} + \frac{1}{\rho^2}\right)n^{-1}$.
\end{theorem}

\begin{Proof}
From (\ref{ring_recursion}), we write the freshness of a node $F_{S_1}$ as 
\begin{align} \label{Fs1_alternative_ring}
   F_{S_1}=  \sum_{i=1}^{n-1} a_i^{(n)} +\frac{n}{n-1}a_{n-1}^{(n)} F_{S_n},
\end{align}
where $a_i^{(n)}$ is given for $i =1,\dots,n-1$ as
\begin{align} 
a_i^{(n)} & = \frac{i}{n} \prod_{j=1}^{i} \frac{1}{1+\rho+\frac{j}{n}} \\
 & =\frac{i}{n} \frac{1}{(1+\rho)^i} \prod_{j=1}^{i} \frac{1}{1+ \frac{1}{1+\rho} \frac{j}{n}}. \label{eqn:a_i} 
\end{align}
From \cite[eqns.~(5)-(6)]{Buyukates21c}, the product term in (\ref{eqn:a_i}) can be approximated as $e^{-\frac{i^2}{2(1+\rho)n}}$, and $a_i^{(n)}$ in (\ref{eqn:a_i}) can be approximated as $a_i^{(n)} \approx \frac{i}{n}\frac{1}{(1+\rho)^i} e^{-\frac{i^2}{2(1+\rho)n}} \approx \frac{i}{n}\frac{1}{(1+\rho)^i}$. Then, since $\sum_{i=1}^{\infty} i \frac{1}{(1+\rho)^i} = \frac{1}{\rho} +\frac{1}{\rho^2}$, and the second term in (\ref{Fs1_alternative_ring}) is negligible, we have $F_{S_1} \approx \sum_{i=1}^{n-1} a_i^{(n)}\approx \left(\frac{1}{\rho} +\frac{1}{\rho^2}\right)\frac{1}{n}$. 
\end{Proof}

\subsection{Fully Connected Networks}
In this network, the end-nodes are fully connected, and each end-node is updated by each of the remaining end-nodes with rate $\frac{\lambda}{n-1}$ and by the source with rate $\frac{\lambda}{n}$. Let $S_j$ denote a subset of any arbitrary $j$ nodes. Since $\lambda_s(S_j) =\frac{j\lambda}{n} $ and $N(S_j) = \mathcal{N}\setminus S_j$. Thus, from (\ref{freshness_gossip_res}), we have 
\begin{align}\label{eqn:fully_connected}
    F_{S_j} = \frac{\frac{j\lambda}{n}+ \frac{j(n-j)\lambda}{n-1}F_{S_{j+1}}}{\lambda_e +\frac{j\lambda}{n}+\frac{j(n-j)\lambda}{n-1}},
\end{align}
for $j =1, \dots, n-1,$ and $F_{S_n} = \frac{1}{1+ \rho}$. Theorem~\ref{fully_connected_thm} below gives the freshness of a node for large $n$. We note that the freshness in a fully connected network in Theorem~\ref{fully_connected_thm} is larger than the freshness in disconnected or ring networks in Theorems~\ref{disconnected_thm} and \ref{ring_thm}. Thus, connectedness improves freshness in gossip networks.

\begin{theorem}\label{fully_connected_thm}
  In a fully connected network, the average freshness of a single node decreases down to $0$ as $n^{-\rho}$ when $0<\rho<1$; as $\frac{\log (n)}{n}$ when $\rho=1$; and as $n^{-1}$ when $\rho>1$.
\end{theorem}

\begin{Proof}
Using (\ref{eqn:fully_connected}), we write the freshness of a node $F_{S_1}$ as 
\begin{align} \label{Fs1_alternative}
   F_{S_1}= \sum_{i=1}^{n-1} b_i^{(n)} +\frac{n}{n-1} b_{n-1}^{(n)} F_{S_n},
\end{align}
where $b_i^{(n)}$ is given for $i=1,\dots,n-1$ as
\begin{align} \label{eqn:b_i}
   b_i^{(n)} \!=\!\frac{1}{1+\frac{n \rho}{i} + \frac{n}{n-1}(n-i)} \prod_{j=1}^{i-1} \frac{1}{1\!+\!\frac{(n-1)\rho}{(n-j)j}\!+\!\frac{n-1}{n}\frac{1}{n-j}}. 
\end{align}
Then, we have
\begin{align} \label{eqn_log_b_i}
    \!\!\!\!-\log (b_i^{(n)}) \!=& \log \left(1+\frac{n\rho}{i}+\frac{n}{n-1}(n-i)\right) \nonumber \\&+ \! \sum_{j=1}^{i-1} \log \left(1\!+\!\frac{(n-1)\rho}{(n-j)j} +\frac{n-1}{n}\frac{1}{n-j} \right). 
\end{align}
Inside the second $\log(\cdot)$ in (\ref{eqn_log_b_i}), $\frac{(n-1)\rho}{(n-j)j}=\frac{(n-1)\rho}{n} \left(\frac{1}{j}+\frac{1}{n-j}\right)$. For large $n$, we have $\frac{n-1}{n}\approx 1$, and for small $x$, we have $\log (1+x)\approx x$, then the second term in (\ref{eqn_log_b_i}) becomes $\sum_{j=1}^{i-1} \log \left(1\!+\!\frac{(n-1)\rho}{(n-j)j} +\frac{n-1}{n}\frac{1}{n-j} \right) \approx (1+\rho) \sum_{j=1}^{i-1}\frac{1}{n-j} + \rho \sum_{j=1}^{i-1} \frac{1}{j}.$ For large $i$,  $\sum_{j=1}^{i-1}\frac{1}{j}\approx \log i $ and $\sum_{j=1}^{i-1}\frac{1}{n-j}\approx \log n - \log (n-i)$. Thus, $-\log(b_i^{(n)})$ in (\ref{eqn_log_b_i}) becomes
\begin{align}\label{eqn_log_b_i_app}
  \!\!\!\!-\log(b_i^{(n)})  =c + (1+\rho) \log n 
  -\rho \log (n-i)+\rho \log i,
\end{align}
where $c = \log \left(1+\frac{\rho}{i}+\frac{1+\rho}{n-i}\right)$, and $\log (1) \leq c\leq \log(2+\rho)$. Thus, we rewrite $b_i^{(n)} $ in (\ref{eqn:b_i}) as $b_i^{(n)} = \frac{d}{n} \left(\frac{1}{i}-\frac{1}{n}\right)^\rho$ where $d = e^{-c}$, which is a constant between 1 and $1/(2+\rho)$. Since the last term in (\ref{Fs1_alternative}) is negligible for large $n$, we have 
\begin{align} \label{Fs1_alternative2}
   F_{S_1}= \sum_{i=1}^{n-1} b_i^{(n)} = \frac{d}{n} \sum_{i=1}^{n-1} \left(\frac{1}{i}-\frac{1}{n}\right)^\rho.
\end{align}
By noting that $\frac{1}{2^\rho}\frac{1}{i^\rho} \leq \left(\frac{1}{i}-\frac{1}{n}\right)^\rho$ for $1\leq i\leq \frac{n}{2}$, we have 
\begin{align}
    \frac{1}{2^\rho}\sum_{i=1}^{n/2}\frac{1}{i^\rho} \leq \sum_{i=1}^{n-1} \left(\frac{1}{i}-\frac{1}{n}\right)^\rho \leq \sum_{i=1}^{n-1}\frac{1}{i^\rho}.
\end{align}
By using Riemann sum, for large $n$, $\sum_{i=1}^{n}\frac{1}{i^\rho}$ is equal to 
\begin{align} \label{rieman_sum}
    \sum_{i=1}^{n}\frac{1}{i^\rho} = \begin{cases} 
      \frac{n^{1-\rho}}{1-\rho}, & \text{when } 0<\rho < 1, \\
      \log n, & \text{when } \rho = 1, \\
      \text{constant}, & \text{when } \rho  >  1 .
   \end{cases}
\end{align}
Thus, $F_{S_1}$ in (\ref{Fs1_alternative2}) becomes
\begin{align} \label{Fs1_alternative3}
   F_{S_1}\!\approx \!\frac{d}{n} \sum_{i=1}^{n-1} \left(\frac{1}{i}-\frac{1}{n}\right)^\rho \!\approx\! \begin{cases} 
      \frac{1}{n^\rho}, & \text{when }0< \rho < 1, \\
      \frac{\log n}{n}, & \text{when } \rho = 1, \\
      \frac{1}{n}, & \text{when } \rho  >  1 ,
   \end{cases}
\end{align}
which completes the proof.
\end{Proof}

\section{Freshness in Clustered Gossip Networks}\label{sect:comm_age}
In this section, similar to \cite{Buyukates21c}, we consider clustered networks, in which each cluster has an associated cluster head that receives updates from the source and forwards them to the corresponding nodes in its own cluster. The total of $n$ nodes are divided into $m$ clusters, with $k$ nodes in each cluster, thus $n=km$. Here, the source updates the cluster heads with a total update rate of $\lambda_s$ (thus, $\frac{\lambda_s}{m}$ update rate per cluster head); each cluster head updates its end-nodes with a total update rate of $\lambda_c$ (thus, $\frac{\lambda_c}{k}$ update rate per end-node); and each end-node updates its immediate neighbors with a total update rate of $\lambda$. Since the nodes (among themselves) and the cluster heads (among themselves) are symmetrical, freshness experienced by each end-node is statistically identical. Thus, in the following analysis, we consider the freshness of a typical node in an arbitrary cluster. In addition to the aforementioned definitions of $\lambda_i(S)$ for $i=\{1,\ldots,n\}$ and $N(S)$ for an arbitrary subset $S$ of the nodes, in what follows, $\lambda_{c}(S)$ denotes the total update rate of a cluster head into the subset $S$.

In Theorem~\ref{thm_freshness_set}, we characterize the freshness in clustered gossip networks.

\begin{theorem}\label{thm_freshness_set}
In a clustered network with $m$ clusters of $k$ nodes each, the freshness of a subset $S$ is given by 
\begin{align}
    F_S = \frac{\lambda_c(S)F_c+\sum_{i \in N(S)} \lambda_i(S) F_{S\cup \{i\}} }{ \lambda_e+\lambda_c(S)+\sum_{i \in N(S)} \lambda_i(S)},\label{freshness_recursion_general}
\end{align}
with ${F}_{c} = \frac{\lambda_s}{\lambda_s + m \lambda_e}$. 
\end{theorem}

The proof of Theorem~\ref{thm_freshness_set} follows from the application of Theorem~\ref{freshness_gossip} to clustered networks. In the following subsections, we characterize the information freshness recursions for the disconnected, ring and fully connected cluster models.   

\subsection{Disconnected Clusters}
The overall disconnected clustered network behaves like a two hop multicast network. In the first hop, the source sends the updates to $m$ cluster heads. In the second hop, each cluster head sends updates to $k$ nodes within its cluster. Different from the multicast network studies in \cite{Zhong17a, Zhong18b, Buyukates18, Buyukates18b, Buyukates19}, here, we consider freshness of information at the end-nodes by random gossiping without applying any central control over the update flows. 

Let $S_j$ denote an arbitrary subset of $j$ nodes in a cluster. Since nodes are disconnected, we have $N(S_1) =\emptyset$ and thus, we write the freshness of a single node as
\begin{align}\label{frehness_disconnected}
    F_{S_1} = \frac{\lambda_c F_c}{\lambda_c+k\lambda_e} = \frac{\lambda_c}{\lambda_c+k\lambda_e} \frac{\lambda_s}{\lambda_s+m\lambda_e}.
\end{align}
In this network, a single node is serially connected to the source via its cluster head. Thus, the freshness in (\ref{frehness_disconnected}) has the same form as in (\ref{example1}). Next, we consider how fast the freshness goes down to $0$ as the number of nodes, $n$, grows large.  

\begin{theorem}\label{theorem_disconn}
 In a clustered disconnected network, the freshness of a node decreases down to $0$ as $\frac{\lambda_c \lambda_s}{\lambda_e^2}n^{-1}$.
\end{theorem}

We note that even with the use of clusters, the freshness of a node still decreases down to $0$ with rate $n^{-1}$. When $\lambda_c >\lambda_e$, the convergence rate is slightly improved, i.e., it increases to $\frac{\lambda_c \lambda_s}{\lambda_e^2}n^{-1}$ in Theorem~\ref{theorem_disconn} from $\frac{\lambda_s}{\lambda_e}$ in Theorem~\ref{disconnected_thm}. When $\lambda_c <\lambda_e$, the freshness of a node converges to $0$ faster. That is, the presence of cluster heads may hurt information freshness in a disconnected network when the update rates of cluster heads are not sufficiently large.      

\subsection{Ring Clusters}
With ring clusters, there are two nodes that update the subset $S_j$ with a total rate $\lambda$. From (\ref{freshness_recursion_general}), we write
\begin{align}\label{frehness_ring}
    F_{S_j} = \frac{\frac{j \lambda_c}{k}F_c+\lambda F_{S_{j+1}}}{\lambda_e +\frac{j \lambda_c}{k}+\lambda}, 
\end{align}
for $j = 1,\dots, k-1$, and we have $F_{S_k} = \frac{\lambda_c}{\lambda_c+\lambda_e} \frac{\lambda_s}{\lambda_s+m\lambda_e}$. For $j=k$, the network simply becomes a two hop network where the first hop is from the source to the cluster head and the second hop is from the cluster head to the entire cluster.

\subsection{Fully Connected Clusters}
With fully connected clusters, each node is connected to other $k-1$ nodes with rates $\frac{\lambda}{k-1}$. Let $S_j$ denote an arbitrary $j$-node subset in a cluster. Subset $S_j$ receives updates from the cluster head with rate $\frac{j\lambda_c}{k}$ and from each of the remaining $k-j$ nodes in the same cluster with rate $\frac{\lambda}{k-1}$. With these, from (\ref{freshness_recursion_general}), we write
\begin{align}\label{frehness_fully}
    F_{S_j} = \frac{\frac{j \lambda_c}{k}F_c+\frac{j(k-j)\lambda}{k-1} F_{S_{j+1}}}{\lambda_e +\frac{j \lambda_c}{k}+\frac{j(k-j)\lambda}{k-1}}, 
\end{align}
for $j = 1,\dots, k-1$, and we have $F_{S_k} = \frac{\lambda_c}{\lambda_c+\lambda_e} \frac{\lambda_s}{\lambda_s+m\lambda_e}$. 

In the following section, we provide numerical results for the scaling of freshness in large gossip networks, as well as for the optimal selection of $m$ and $k$ for clustered network models with different $\lambda$, $\lambda_s$, and $\lambda_c$ values. 

\section{Numerical Results}
\subsection{Numerical Results for Large Gossip Networks}
In this subsection, we provide numerical results for the scaling of information freshness in large gossip networks. For all network types, since information freshness decreases to $0$, we consider the scaling of the inverse freshness function of a node, i.e., $F_{S_1}^{-1} = \frac{1}{F_{S_1}}$, and vary $n$ in between $500$ to $100,000$. 

\begin{figure}[t]
 	\begin{center}
 	\subfigure[]{%
 	\includegraphics[width=0.49\linewidth]{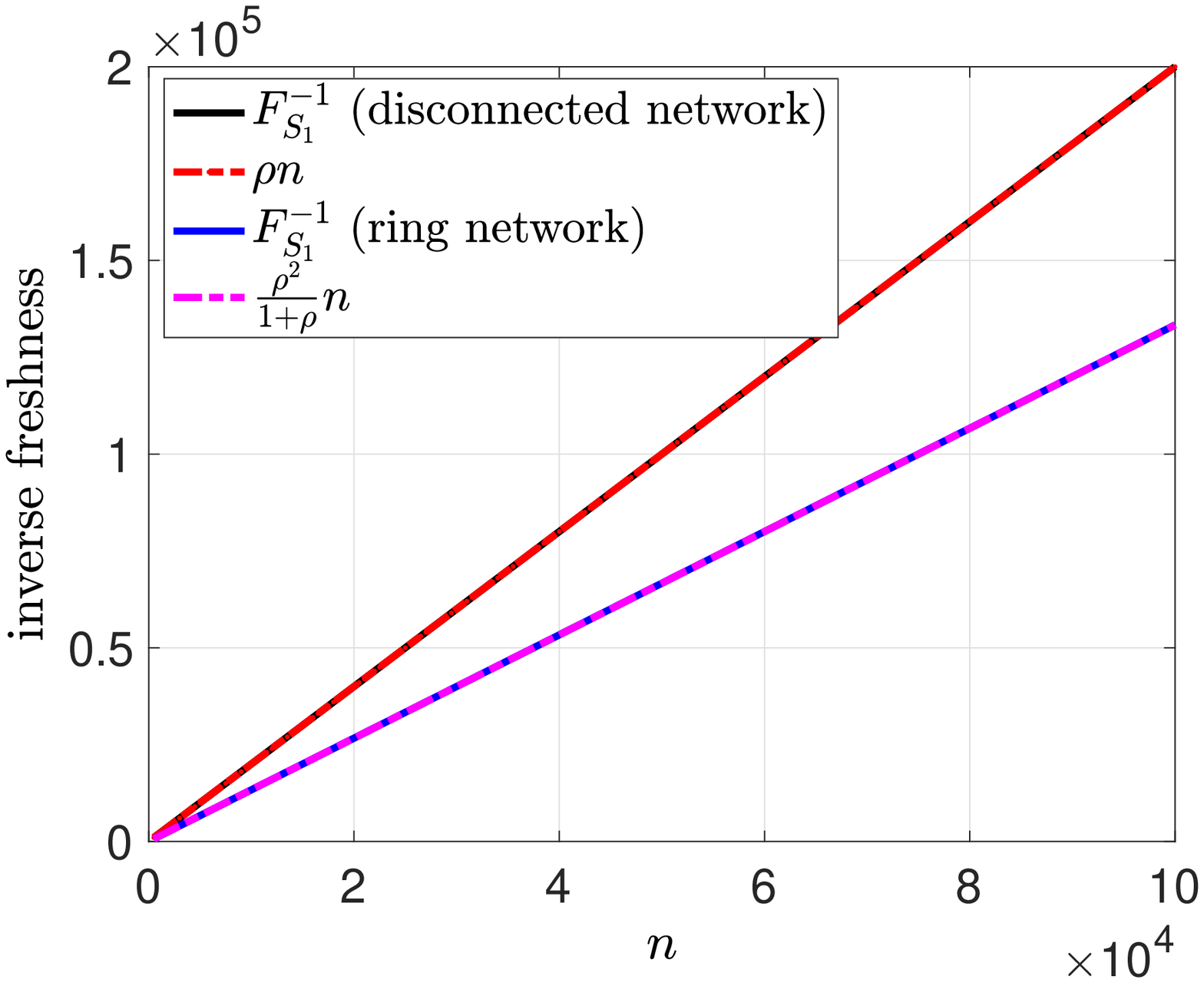}}
 	\subfigure[]{%
 	\includegraphics[width=0.49\linewidth]{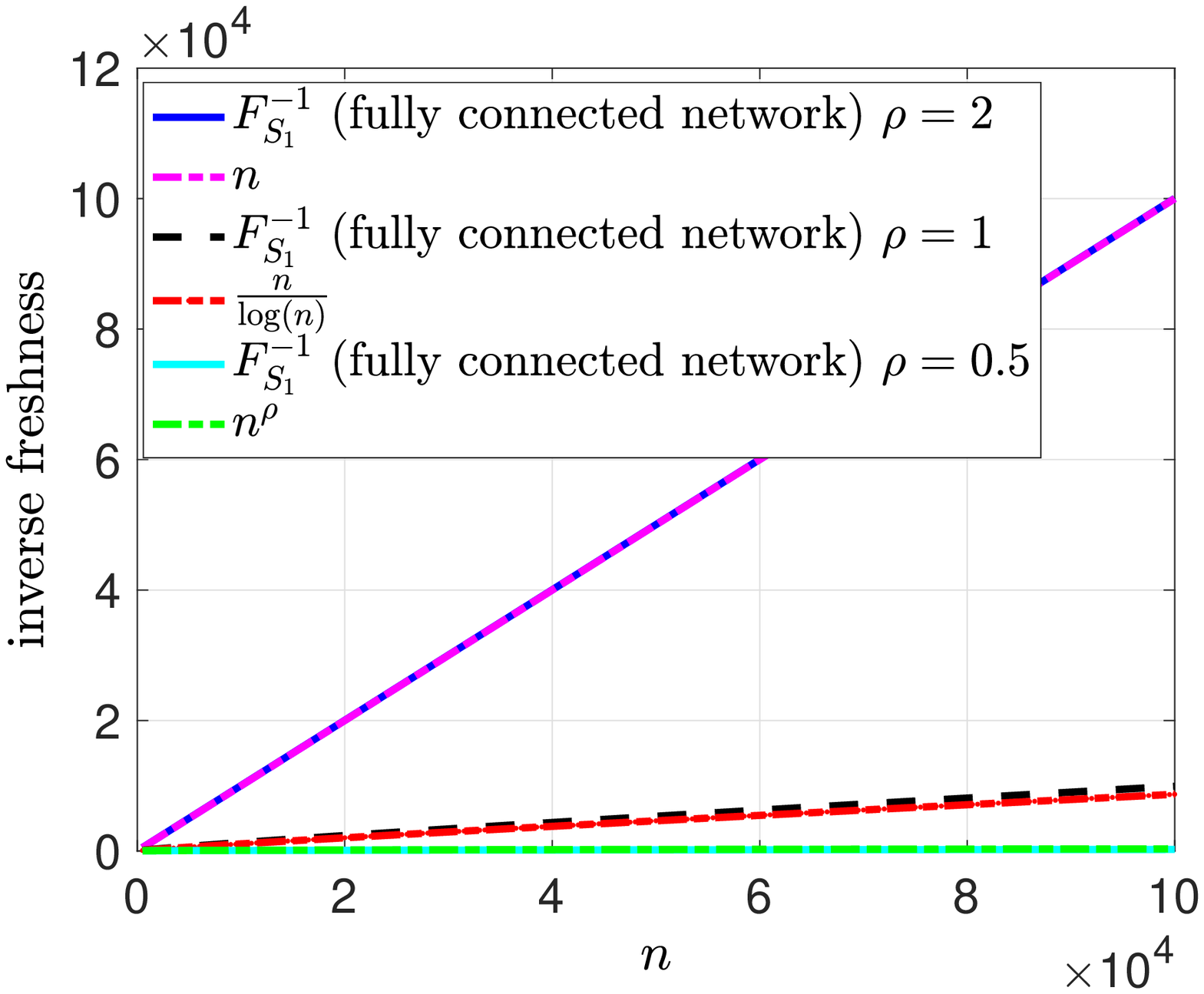}}
 	\end{center}
 	\vspace{-0.5cm}
 	\caption{Scaling of inverse freshness of a node (a) in disconnected and ring networks when $\rho=2$ ($\lambda_e = 2$ and $\lambda =1$), (b) in a fully connected network when $\rho=2$ ($\lambda_e = 2$ and $\lambda =1$), $\rho=1$ ($\lambda_e = 1$ and $\lambda =1$), and $\rho=0.5$ ($\lambda_e = 0.5$ and $\lambda =1$).}
 	\label{Fig:scaling1_2}
 	\vspace{-0.5cm}
 \end{figure}

In the first numerical result, we consider disconnected and ring networks for $\lambda_e=2$, and $\lambda=1$, i.e., $\rho=2$. In Fig.~\ref{Fig:scaling1_2}(a), we see that the scaling of inverse freshness of a node is $F_{S_1}^{-1} \approx \rho n$ for the disconnected network, and $F_{S_1}^{-1} \approx \frac{\rho^2}{1+\rho} n$ for the ring network. Although the inverse freshness increases linearly with $n$ for both network types, we see in Fig.~\ref{Fig:scaling1_2}(a) that the slope of inverse freshness in the ring network is smaller, indicating a slower decrease to $0$ for freshness. This verifies that increased connectedness improves freshness in gossip networks.     

In the second numerical result, we consider a fully connected network when $\rho =2$ ($\lambda_e=2$, and $\lambda=1$), $\rho =1$ ($\lambda_e=1$, and $\lambda=1$), and $\rho =0.5$ ($\lambda_e=0.5$, and $\lambda=1$). In Fig.~\ref{Fig:scaling1_2}(b), we see that when $\rho =2$, i.e., when the update rates of a node and the source (both are $\lambda$) are smaller compared to the information change rate at the source $\lambda_e$, then the inverse freshness still scales linearly with $n$. When $\lambda$ gets large (specifically, when $\lambda =\lambda_e$), inverse freshness of a node starts to scale with $\frac{n}{\log (n)}$ as shown in Fig.~\ref{Fig:scaling1_2}(b). When we further increase $\lambda$, i.e., when $\lambda>\lambda_e$ (or $\rho<1$), the inverse freshness scales with $n^{\rho}$, i.e., sublinearly. Thus, when the update rates of the source and the end-nodes are large enough in the fully connected network, freshness can be improved from $\frac{1}{n}$ to $\frac{1}{n^{\rho}}$.            

\subsection{Numerical Results for Clustered Gossip Networks}
In this subsection, we provide numerical results regarding the optimal cluster size selection $k$ in clustered networks for the cases of disconnected, ring and fully connected clusters, for different update rates at the source, cluster heads and the nodes, when information change rate at the source is $\lambda_e=1$ and the number of nodes is $n=120$.

\begin{figure}[t]
 	\begin{center}
 	\subfigure[]{%
 	\includegraphics[width=0.49\linewidth]{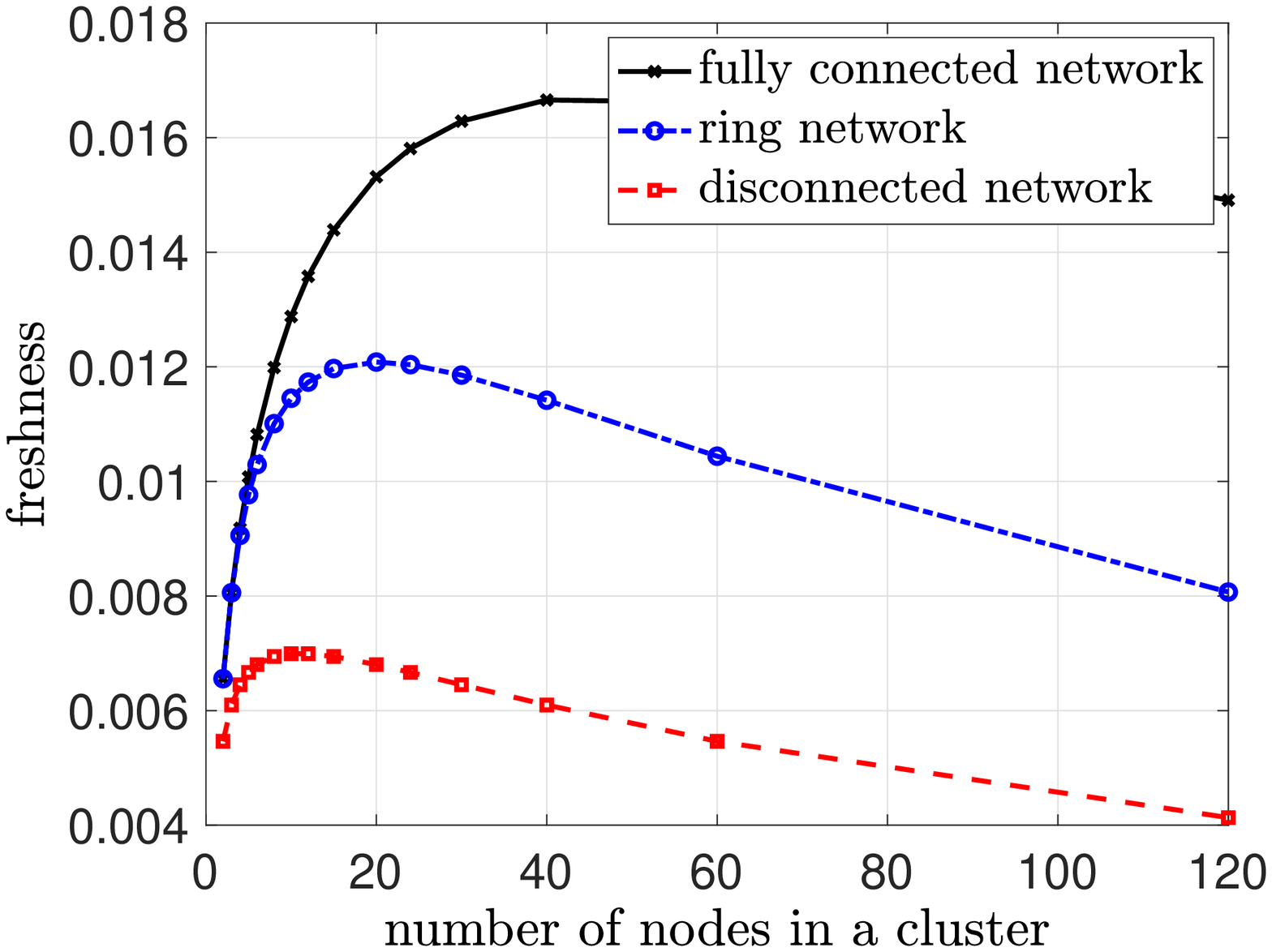}}
 	\subfigure[]{%
 	\includegraphics[width=0.49\linewidth]{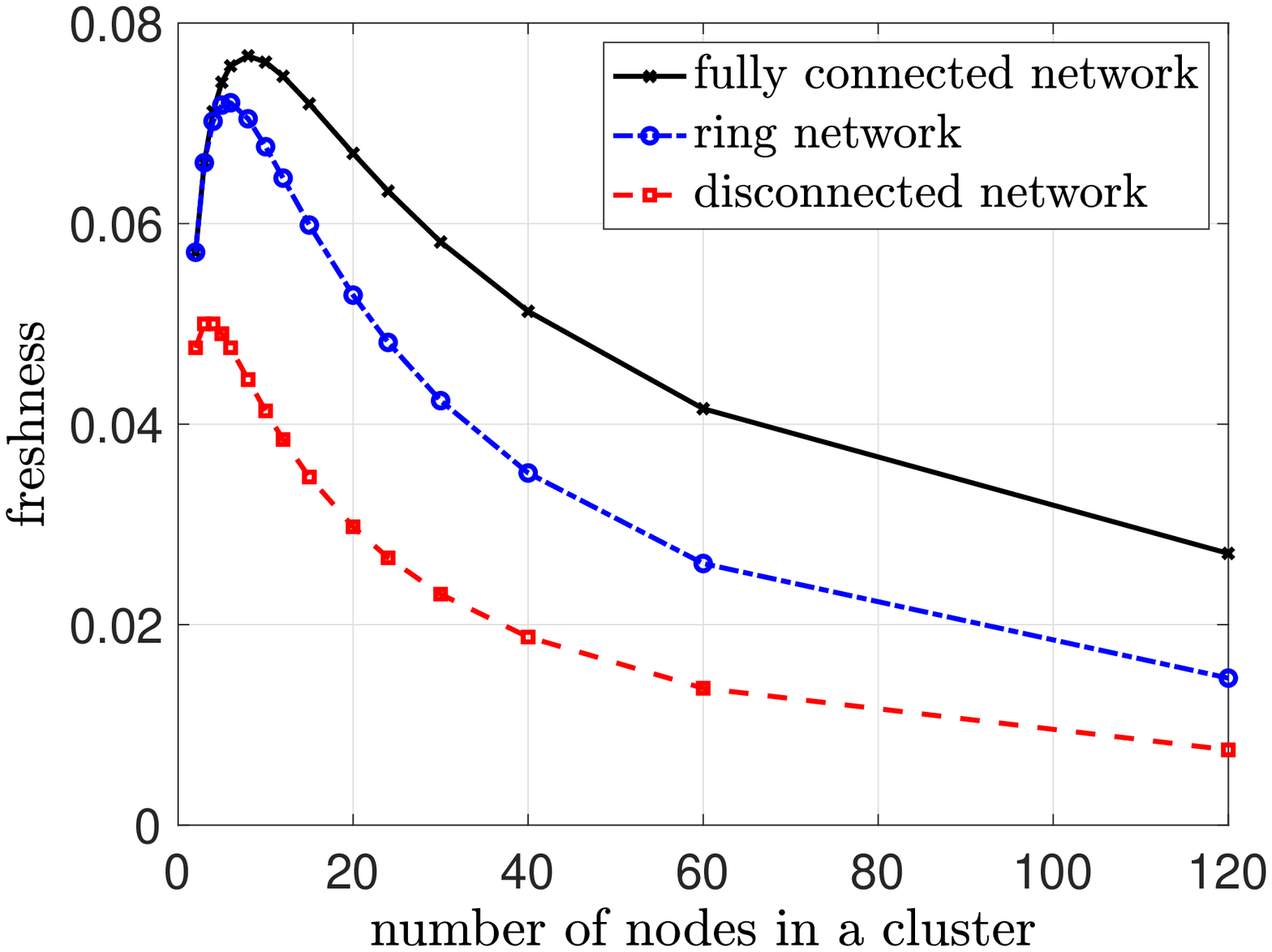}}\\ \vspace{-0.35cm}
 	\subfigure[]{%
 	\includegraphics[width=0.49\linewidth]{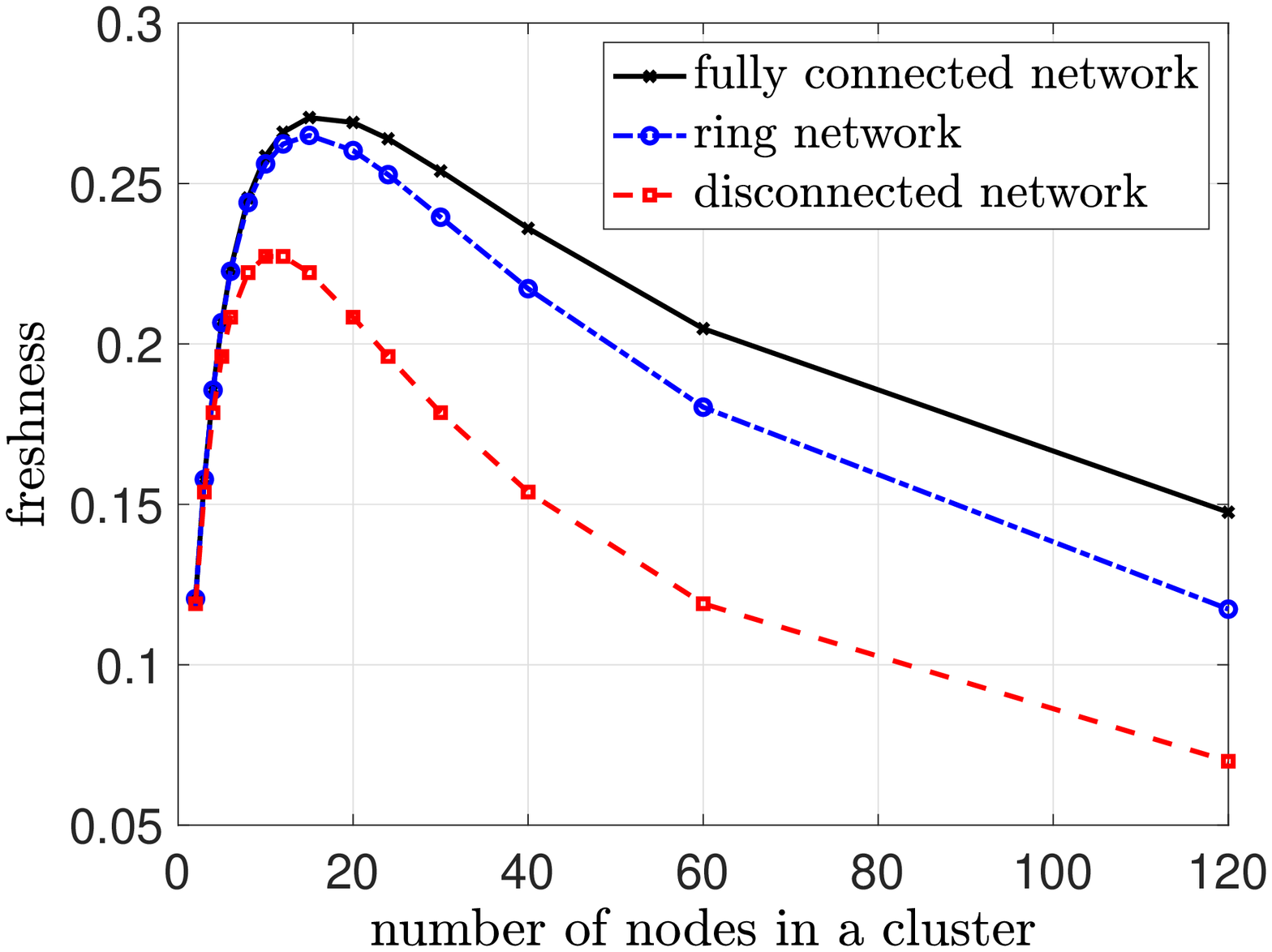}}
 	\subfigure[]{%
 	\includegraphics[width=0.49\linewidth]{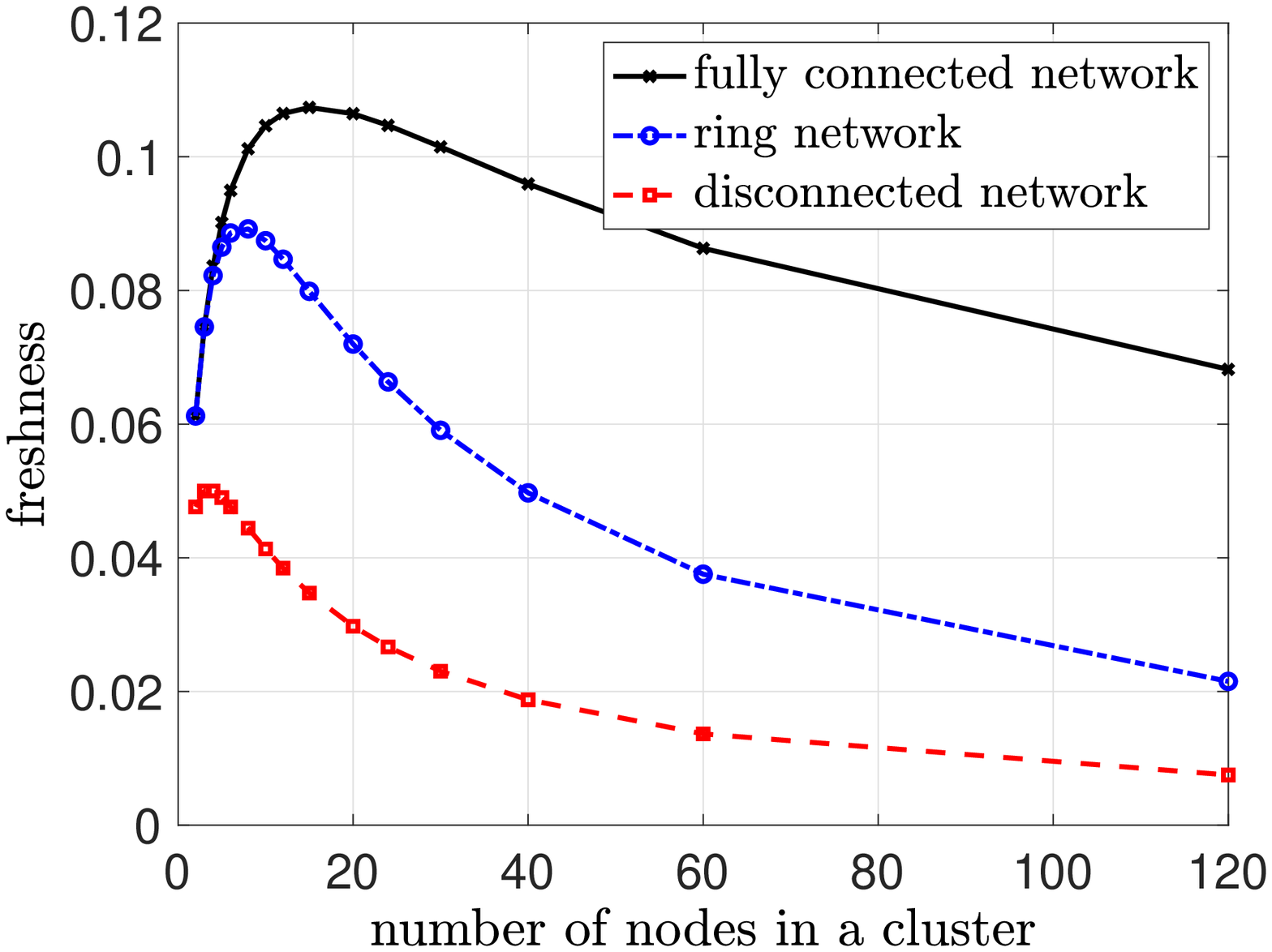}}
 	\end{center}
 	\vspace{-0.4cm}
 	\caption{Binary freshness of a node with disconnected, ring and fully connected clusters with $n=120$, $\lambda_e =1$, (a) $\lambda_{s} =1$, $\lambda_{c} =1$, $\lambda =1$, (b) $\lambda_{s} =10$, $\lambda_{c} =1$, $\lambda =1$, (c) $\lambda_{s} =10$, $\lambda_{c} =10$, $\lambda =1$, (d) $\lambda_{s} =10$, $\lambda_{c} =1$, $\lambda =2$. }
 	\label{Fig:sim_results_all_v2}
 	\vspace{-0.5cm}
 \end{figure}

First, we take $\lambda_{s} =1$, $\lambda_{c} =1$, $\lambda =1$. In Fig.~\ref{Fig:sim_results_all_v2}(a), we see that the optimal cluster size is $k^* = 40$ in fully connected clusters; $k^* = 20$ in ring clusters; $k^* = 10$ or $k^* = 12$ in disconnected clusters. We observe that the optimal cluster size increases as the connectivity among the nodes increases. Further, the achievable freshness increases with the connectivity within the clusters.  

Second, we consider $\lambda_{s} =10$, $\lambda_{c} =1$, $\lambda =1$, i.e., the update rate of the source is increased compared to the setting in Fig.~\ref{Fig:sim_results_all_v2}(a). In Fig.~\ref{Fig:sim_results_all_v2}(b), we see the trade-off between the number of clusters and the number of nodes in a cluster. Even though we increase the update rate of the source, since it is not large enough, freshness initially increases with the cluster size $k$ as the total number of clusters decreases. That is, the source can send updates to each cluster more efficiently with increasing $k$. On the other hand, since the update rates at the cluster heads are also limited, further increasing $k$ starts to decrease the freshness at the end-nodes. Thus, there is a sweet spot where the freshness is maximized for each cluster topology. Since we increase $\lambda_{s}$ compared to the previous example, the source can support more cluster heads and the optimal cluster sizes become $k^*= 8$ for  fully connected clusters, $k^*= 6$ for ring clusters, $k^*= 3$ or $k^*= 4$ for disconnected clusters.

Third, we consider $\lambda_{s} =10$, $\lambda_{c} =10$, and $\lambda =1$, i.e., we increase the cluster heads' update rate compared to the second example. Since we increase the update rate of the cluster heads, the optimal cluster size increases from $k^*= 8$ to $k^*= 15$ for fully connected clusters; from $k^*= 6$ to $k^*= 15$ for ring clusters; and from $k^*= 3$ or $k^*= 4$ to $k^*= 10$ or $k^*= 12$ for disconnected clusters as shown in Fig.~\ref{Fig:sim_results_all_v2}(c).

Fourth, we take $\lambda_{s} =10$, $\lambda_{c} =1$, and $\lambda =2$, i.e., compared to the second example, we increase the update rate among the end-nodes. We see in Fig.~\ref{Fig:sim_results_all_v2}(d) that since there is no updates among the nodes in the disconnected clustered network, the optimal cluster size stays the same as $k^*= 3$ or $k^*= 4$. On the other hand, since the update rates among the nodes are increased, the optimal cluster sizes become to $k^*= 15$ for fully connected clusters, and $k^*= 8$ for ring clusters.

\section{Conclusion}
By using the SHS method, we developed recursive formulas to find the binary freshness for a given node in arbitrarily connected networks. Next, we considered structured large gossip networks. We showed that the binary freshness decreases to $0$ with rate $n^{-1}$ both for disconnected and ring networks, but with a strictly slower rate in ring networks. We also showed that binary freshness decreases to $0$ with much slower rates for fully connected networks; in particular, the form of the freshness function changes from $n^{-1}$ to $n^{-\rho}$ when we go to fully connected networks when the update rate of the source is sufficiently large. We also found recursive expressions for the binary freshness in clustered gossip networks. Via numerical results, we studied the effects of the update rates of the source, cluster heads and the end-nodes on the optimum cluster size. An interesting future direction is to investigate potential performance gains due to gossiping among the cluster heads.

\bibliographystyle{unsrt}
\bibliography{IEEEabrv,lib_v6}

\begin{thebibliography}{10}

\bibitem{Kaul12a}
S.~K. Kaul, R.~D. Yates, and M.~Gruteser.
\newblock Real-time status: How often should one update?
\newblock In {\em IEEE Infocom}, March 2012.

\bibitem{YatesSurvey}
R.~D. Yates, Y.~Sun, R.~Brown, S.~K. Kaul, E.~Modiano, and S.~Ulukus.
\newblock Age of information: An introduction and survey.
\newblock {\em IEEE Journal on Selected Areas in Communications},
  39(5):1183--1210, May 2021.

\bibitem{cho03}
J.~Cho and H.~Garcia-Molina.
\newblock Effective page refresh policies for web crawlers.
\newblock {\em ACM Transactions on Database Systems}, 28(4):390--426, December
  2003.

\bibitem{kolobov19a}
A.~Kolobov, Y.~Peres, E.~Lubetzky, and E.~Horvitz.
\newblock Optimal freshness crawl under politeness constraints.
\newblock In {\em ACM SIGIR Conference}, pages 495--504, July 2019.

\bibitem{Bastopcu2021}
M.~{Bastopcu} and S.~{Ulukus}.
\newblock Information freshness in cache updating systems.
\newblock {\em IEEE Transactions on Wireless Communications}, 20(3):1861--1874,
  March 2021.

\bibitem{Bastopcu20e}
M.~Bastopcu and S.~Ulukus.
\newblock Maximizing information freshness in caching systems with limited
  cache storage capacity.
\newblock In {\em Asilomar Conference}, November 2020.

\bibitem{Kaswan21}
P.~Kaswan, M.~Bastopcu, and S.~Ulukus.
\newblock Freshness based cache updating in parallel relay networks.
\newblock In {\em IEEE ISIT}, July 2021.

\bibitem{Bastopcu20c}
M.~Bastopcu and S.~Ulukus.
\newblock Who should {Google} {Scholar} update more often?
\newblock In {\em IEEE Infocom}, July 2020.

\bibitem{Yates21}
R.~D. Yates.
\newblock The age of gossip in networks.
\newblock In {\em IEEE ISIT}, July 2021.

\bibitem{Buyukates21c}
B.~Buyukates, M.~Bastopcu, and S.~Ulukus.
\newblock Age of gossip in networks with community structure.
\newblock In {\em IEEE SPAWC}, September 2021.

\bibitem{Eryilmaz21}
B.~Abolhassani, J.~Tadrous, A.~Eryilmaz, and E.~Yeh.
\newblock Fresh caching for dynamic content.
\newblock In {\em IEEE Infocom}, May 2021.

\bibitem{Maatouk20b}
A.~Maatouk, S.~Kriouile, M.~Assaad, and A.~Ephremides.
\newblock The age of incorrect information: A new performance metric for status
  updates.
\newblock {\em IEEE/ACM Transactions on Networking}, 28(5):2215--2228, July
  2020.

\bibitem{Maatouk20a}
A.~{Maatouk}, M.~{Assaad}, and A.~{Ephremides}.
\newblock The age of incorrect information: an enabler of semantics-empowered
  communication.
\newblock December 2020.
\newblock Available on arXiv:2012.13214.

\bibitem{Kam20a}
C.~Kam, S.~Kompella, and A.~Ephremides.
\newblock Age of incorrect information for remote estimation of a binary markov
  source.
\newblock In {\em IEEE Infocom}, July 2020.

\bibitem{Zhong18c}
J.~Zhong, R.~D. Yates, and E.~Soljanin.
\newblock Two freshness metrics for local cache refresh.
\newblock In {\em IEEE ISIT}, June 2018.

\bibitem{Zhong17a}
J.~Zhong, E.~Soljanin, and R.~D. Yates.
\newblock Status updates through multicast networks.
\newblock In {\em Allerton Conference}, October 2017.

\bibitem{Zhong18b}
J.~Zhong, R.~D. Yates, and E.~Soljanin.
\newblock Multicast with prioritized delivery: How fresh is your data?
\newblock In {\em IEEE SPAWC}, June 2018.

\bibitem{Buyukates18}
B.~Buyukates, A.~Soysal, and S.~Ulukus.
\newblock Age of information in two-hop multicast networks.
\newblock In {\em Asilomar Conference}, October 2018.

\bibitem{Buyukates18b}
B.~Buyukates, A.~Soysal, and S.~Ulukus.
\newblock Age of information in multihop multicast networks.
\newblock {\em Journal of Communications and Networks}, 21(3):256--267, July
  2019.

\bibitem{Buyukates19}
B.~Buyukates, A.~Soysal, and S.~Ulukus.
\newblock Age of information in multicast networks with multiple update
  streams.
\newblock In {\em Asilomar Conference}, November 2019.

\end{thebibliography}
\end{document}